\documentclass[article]{cjaa}

\newcommand\arcdeg{\mbox{$^\circ$}}%
\newcommand\Smin{S_{\rm min}}
\newcommand\Dmax{D_{\rm max}}
\newcommand\gtrsim{\ga}%
\newcommand\lesssim{\la}%

\usepackage{graphicx}                   
\usepackage{natbib}
\bibpunct{(}{)}{;}{a}{}{,}

\begin{document}

\title{Arecibo and the ALFA Pulsar Survey}

   \volnopage{Vol.0 (200x) No.0, 000--000}      
   \setcounter{page}{1}          

\author{
  J.~van~Leeuwen \inst{1}
  \and J.~M.~Cordes     
  \and D.~R.~Lorimer    
  \and P.~C.~C.~Freire  
  \and F.~Camilo        
  \and I.~H.~Stairs     
  \and D.~J.~Nice       
  \and D.~J.~Champion   
  \and R.~Ramachandran  
  \and A.~J.~Faulkner   
  \and A.~G.~Lyne       
  \and S.~M.~Ransom     
  \and Z.~Arzoumanian   
  \and R.~N.~Manchester 
  \and M.~A.~McLaughlin 
  \and J.~W.~T.~Hessels 
  \and W.~Vlemmings     
  \and A.~A.~Deshpande  
  \and N.~D.~R.~Bhat    
  \and S.~Chatterjee    
  \and J.~L.~Han        
  \and B.~M.~Gaensler   
  \and L.~Kasian        
  \and J.~S.~Deneva     
  \and B.~Reid          
  \and T.~J.~W.~Lazio   
  \and V.~M.~Kaspi      
  \and F.~Crawford      
  \and A.~N.~Lommen     
  \and D.~C.~Backer     
  \and M.~Kramer        
  \and B.~W.~Stappers   
  \and G.~B.~Hobbs      
  \and A.~Possenti      
  \and N.~D'Amico       
  \and C.-A.~Faucher-Gigu\`ere
  \and M.~Burgay
}

\institute{
  Department of Physics \& Astronomy, University of British Columbia,
  6224 Agricultural Road, Vancouver B.C. V6T 1Z1, Canada, 
  {\tt joeri@astro.ubc.ca}
\and 
  {\tt http://alfa.naic.edu/pulsars}
}
   \date{Received~~2005; accepted~~2005}

\abstract{ The recently started Arecibo L-band Feed Array (ALFA)
pulsar survey aims to find $\sim 1000$ new pulsars. Due 
to its high time and frequency resolution the survey is especially
sensitive to millisecond pulsars, which have the potential to test
gravitational theories, detect gravitational waves and probe the
neutron-star equation of state.  Here we report the results of our
preliminary analysis: in the first months we have discovered 21 new
pulsars. One of these, PSR~J1906+0746, is a young 144-ms pulsar in a
highly relativistic 3.98-hr low-eccentricity orbit. The
$2.61\pm0.02$~$M_{\odot}$ system is expected to coalesce in $\sim
300$~Myr and  contributes significantly to the computed
cosmic inspiral rate of compact binary systems.
}
\authorrunning{van Leeuwen et al.}
\maketitle
\section{Introduction}\label{sec:intro}
Radio pulsars continue to provide unique opportunities for testing
theories of gravity and probing states of matter otherwise
inaccessible \cite{sta03}, and in large samples they allow
detailed modeling of the Galactic neutron star population
\citep{bwhv92, acc02}.

For these reasons, we have initiated a large-scale pulsar survey
\cite[][ hereafter Paper I]{cfl+05} that aims to discover pulsars in
short-period relativistic orbits to test gravitational theories in the
strong-field regime, and millisecond pulsars (MSPs) with ultrastable
spin rates that can be used as detectors of long-period ($>$
years) gravitational waves \cite[e.g.][]{lb01}.  Furthermore,
long-period ($P > 5$~s) and strongly magnetised pulsars may
clarify the connection with magnetars, and the nature of the elusive
radio emission mechanism. Additionally, determining the pulsar
velocity distribution will help constrain aspects of the formation of
neutron stars in core-collapse supernovae \citep[e.g.][]{wkh04}.

The new survey is enabled by several innovations. First is the Arecibo
L-band Feed Array (ALFA), a seven-beam feed and receiver system
designed for large-scale surveys in the 1.2--1.5 GHz band.  The
1.4~GHz operating frequency of ALFA is particularly well suited for
pulsar searching of the Galactic plane.  Lower frequencies suffer the
deleterious effects of pulse broadening from interstellar scattering
and dispersion,
while pulsar flux densities typically are much reduced at higher
frequencies.  The ALFA frontend is similar to the 13-beam system used
for the extremely prolific Parkes multibeam (PMB) pulsar survey of the
Galactic plane \citep{mlc+01}. Our survey will complement the PMB
survey in its sky coverage.

Second, our initial and next-generation spectrometer systems have much
finer resolution in both time and frequency than the spectrometer used
with the PMB, increasing the detection volume of MSPs by an order of
magnitude. Additionally, the sensitivity of the Arecibo telescope
allows for short pointings that simplify the detection of binary
pulsars undergoing strong acceleration.

This sensitivity for relativistic binaries is illustrated by our
discovery of the latest binary system PSR~J1906+0746 \citep[][ hereafter
Paper II]{lsf+06}. Since binaries like PSR~J1906+0746 will coalesce
due to gravitational wave emission well within a Hubble time, their
merger rate \citep[e.g.][]{phi91,kklw04} is of great interest to
the gravitational wave detector community as potential sources for
current interferometers such as LIGO \citep{aad+92}.

Following a brief description of the ALFA survey setup in
\S~\ref{sec:survey} we describe general survey results in
\S~\ref{sec:results} and our discovery of young, highly relativistic
binary pulsar PSR~J1906+0746 in \S~\ref{sec:1906}.  We conclude in
\S~\ref{sec:future}, and outline our plans for the future.

        \begin{figure}[t]%
        \begin{minipage}[b]{70mm}%
          \includegraphics[width=70mm]{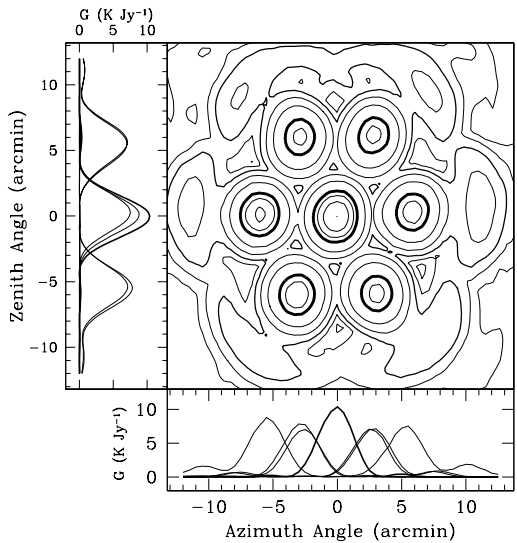}%
          \vspace{-20mm}%
        \end{minipage}%
        \hspace{2mm}%
        \begin{minipage}[b]{70mm}%
          \caption{
          \label{fig:beams}%
	  Contours of the ALFA telescope gain. Levels are at --1, --2,
	  --3, --6, --9, --12, --15 and --19 dB from the central
	  peak. The equivalent circular beam width (FWHM), averaged
	  over all beams, is 3.35 arcmin at 1.42 GHz.  The on-axis
	  gain is approximately 10.4~K~Jy$^{-1}$ 
	  for the central beam and $\sim
	  8.2$~K~Jy$^{-1}$ for the other six beams. The system
	  temperature looking out of the Galactic plane is $\sim 24$~K.
	  (Paper I; Heiles~2004).  }
          \vspace{5mm}%
        \end{minipage}%
        \end{figure}%

\section{The ALFA Pulsar Survey}\label{sec:survey}
The ALFA system sits in the Gregorian focus of the Arecibo
telescope and provides seven 3-arcmin 8$-$10~K~Jy$^{-1}$ beams on the
sky (Figure~\ref{fig:beams}). Within a year, new polyphase-filter
spectrometers will process the full 300 MHz bandwidth with 1024
spectral channels. Currently the four Wideband Arecibo Pulsar
Processor (WAPP) systems \cite[]{dsh00} record 256 channels every
64~$\mu$s, using the 100~MHz band around 1.42~GHz, as that band shows the
least radio-frequency interference.

        \begin{figure}[b]%
        \begin{minipage}[b]{95mm}%
          \includegraphics[width=95mm]{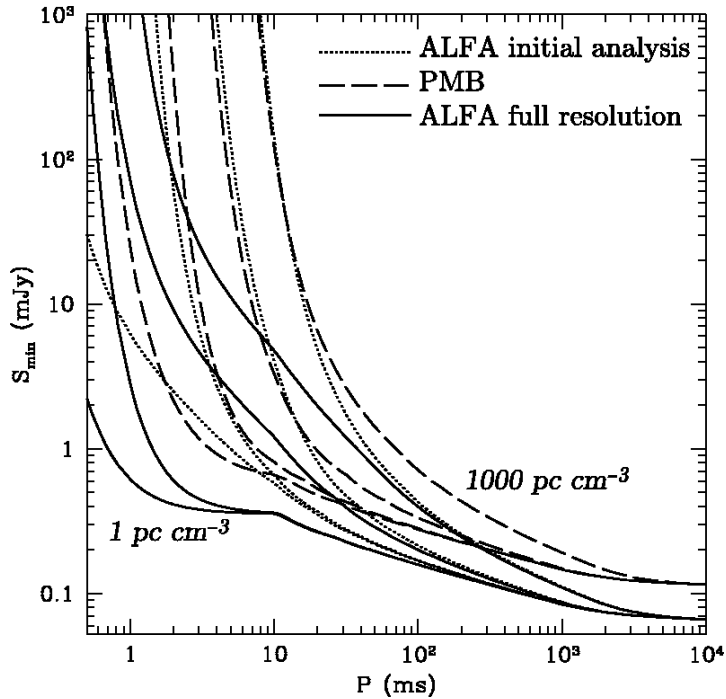}%
        \end{minipage}%
        \hspace{2mm}%
        \begin{minipage}[b]{49mm}%
\caption{\label{fig:smin} Theoretical minimum detectable flux density
($\Smin$) vs. $P$ for different DMs. Real-world effects such as RFI
and receiver gain variations will raise the effective threshold of the
survey over these lower bounds. Per survey, DM values from the lowest
to the highest curve are 1, 200, 500 and 1000~pc~cm$^{-3}$. We assume
the intrinsic pulse duty cycle scales as $P^{-1/2}$ with a maximum of
0.3 at 10 ms, hence the breakpoint at that period (Paper I).  }
          \vspace{5mm}%
        \end{minipage}%
        \end{figure}%

The data are processed twice: during the observations incoming data
are transferred to the Arecibo Signal Processor \cite[]{drb+04}, a
computer cluster that processes the data in quasi-realtime after
reducing the time and frequency resolution to increase
throughput. This analysis, described in detail in Paper I, is
primarily sensitive to pulsars with $P \gtrsim 30$~ms, which are
expected to make up the bulk of all discoveries. In the second
processing step the data will be re-analysed on several computer
clusters at the home institutions of members of the Pulsar ALFA
(PALFA) Consortium. In this step full-resolution data and acceleration
searches will be used, increasing sensitivity to MSPs and pulsars in
short period binary systems as well as to pulsars with large values of
dispersion measure (DM).

With our 134-s integration time, the minimum detectable flux density
$\Smin$ for PALFA is a factor 1.6 smaller than for the PMB survey
(Figure~\ref{fig:smin}), implying a maximum distance $\Dmax \propto
\Smin^{-1/2}$ about 1.3 times larger for long-period pulsars.  The
sampled volume on axis is accordingly about a factor of two larger for
long-period pulsars.  In our full-resolution analysis, the volume
increase is even larger for short periods, owing to the smaller PALFA
channel widths and the shorter sample interval.  For $P\lesssim 10$
ms, the searched volume increase can be a factor of 10 or more (Paper
I).

\section{General Results}\label{sec:results}
\begin{center}
\begin{figure*}[t]
\includegraphics[]{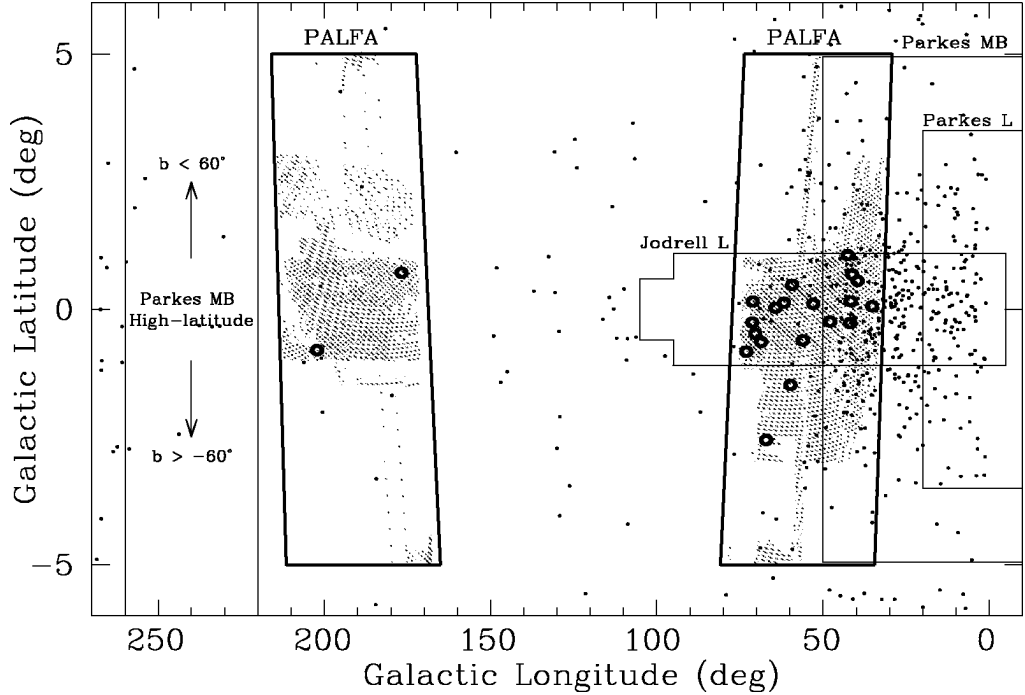}
\caption{ \label{fig:lbregions} 
ALFA surveys regions.  Gray dots indicate pointings so far, small
black dots designate known pulsars and large circles represent newly
discovered pulsars.  We also show boundaries of similar L-band
surveys: the PMB, Parkes \citep{jlm+92}, Jodrell Bank
\citep{clj+92} and Swinburne PMB High-latitude \citep{e+01} surveys.}
\end{figure*}
\end{center}

Between August 2004 and November 2005 we have used 75 hours of
telescope time for 2516 pointings in the Galactic anti-center and 166
hours for 4838 pointings in the inner Galaxy, covering 43~deg$^2$ and
83~deg$^2$ in each region respectively (Figure~\ref{fig:lbregions}).  So
far we have found 21 new pulsars in the first analysis. Seven of these
are in the southernmost region visible from Arecibo. This region was
previously covered by the PMB survey, suggesting that our survey
indeed already surpasses the depth of the PMB survey in the first
analysis.  All previously known pulsars were detected in our pointings
if they were within one beam radius of one of the ALFA beams.  In
addition, we detect some strong pulsars several beam radii from the
nearest beam center.

\section{The young, relativistic binary pulsar J1906+0746}\label{sec:1906}
\subsection{Discovery and follow-up observations}
PSR~J1906+0746 was discovered with a signal-to-noise ratio S/N~$\sim
11$ in data taken on 2004 September 27. The pulsar was 2$\farcm$5
(1.47 beam radii) from the center of the beam, where the antenna gain
is $\sim$5 times smaller than at its center.  PSR~J1906+0746 also lies
in the region of sky covered by the PMB. Examination of the search output of the
35-minute PMB observation of this position showed a 144-ms
periodicity with S/N~$\sim 7$, below the nominal S/N threshold of
8--9. While PSR~J1906+0746 appears with S/N~$\sim25$ in the PMB
``stack-and-slide'' algorithm output \cite{fsk+04}, 
it was not selected as a candidate in the PMB because of
significant amounts of radio-frequency interference close to 144~ms.

The high degree of acceleration detected in the PMB observation
immediately implied that PSR~J1906+0746 is a short-period binary
system. Follow-up observations with the Jodrell Bank, Arecibo, Green
Bank and Parkes telescopes refined the orbital and spin parameters and
the position (Table~\ref{tab:parms}):  PSR~J1906+0746 is a
young object with a characteristic age of 112~kyr and a strong
magnetic field ($B=3.2 \times 10^{19}$). The pulsar is characterized by a
narrow main pulse and significant interpulse feature with high linear
polarization at 1.4~GHz (Figure~\ref{fig:profs}; Paper II).

\begin{table}[b]
  \centering
  \begin{tabular}{ll|ll}%
    \hline%
    Parameter  &   Value & Parameter  &   Value \\
    \hline%
    \hline%
Right ascension (J2000)&$19^{\rm h}06^{\rm m}48\fs673(6)$     &     Projected semi-major axis, $x$ (lt s)          & 1.420198(2)\\	
Declination (J2000)    & $07\arcdeg46\arcmin28.6(3)\arcsec$   &	    Periastron advance rate, $\dot{\omega}$ (\arcdeg~yr$^{-1}$)&7.57(3)\\
Spin period, $P$ (ms)                   & 144.071929982(3)    &     Dispersion measure, DM (cm$^{-3}$\,pc)      & 217.780(2)          \\
Spin period derivative, $\dot P$      & $2.0280(2)\times 10^{-14}$& Rotation measure, RM (rad m$^{-2}$)            & +150(10) \\
Epoch (MJD)                                 & 53590           &     Characteristic age, $\tau_c$ (kyr)     & 112             \\
Orbital period, $P_b$ (days)                & 0.165993045(8)  &	    Total system mass, $M$ ($M_{\odot}$)  &  2.61(2)  \\
Orbital eccentricity, $e$                   & 0.085303(2)     &	    Spin-down power, $\dot E$ (ergs\,s$^{-1}$)& $2.7\times10^{35}$\\
Spectral index, $\alpha$ & --1.3(2)  			      &	    Inferred distance, $d$ (kpc)           & $\sim 5.4$        \\
Mass function, $f$ ($M_{\odot}$)  &  0.1116222(6)             &	    Coalescence time, $\tau_g$ (Myr)& $\sim 300$\\                          
Magnetic field, $B$ (Gauss)         & $1.7\times 10^{12}$     &     Flux density at 1.4\,GHz, $S$ (mJy) & 0.55(15)   \\ \\
\hline%
\end{tabular}
\caption{\label{tab:parms}Observed and derived parameters of 
PSR~J1906+0746 (Paper II).}%
\end{table}

\subsection{Nature of Companion}
\label{sec:nature}

From the orbital parameters of PSR~J1906+0746, we infer a Keplerian
mass function $f(m_p,m_c)=(m_c \sin i)^3/(m_p+m_c)^2 = 0.11 \,
M_{\odot}$. Here $m_p$ is the pulsar mass, $m_c$ is the companion mass
and $i$ is the angle between the orbital plane and the plane of the
sky.  Our measurement of the orbital periastron advance
$\dot{\omega}=7.57\pm0.03~$\arcdeg~yr$^{-1}$, is, after the double
pulsar system \cite{bdp+03}, the second largest observed so far.
Interpreting this large value within the framework of general
relativity implies that the total system mass
$M\,=\,m_p+m_c\,=\,2.61\pm0.02\,M_{\odot}$ (Paper II).  Measured
masses of the neutron stars in double neutron star binary systems
range from 1.25~M$_{\odot}$ \cite{lbk+04} to 1.44~M$_{\odot}$
\cite{wt03}, so it is likely that
the mass of PSR~J1906+0746 is within these limits. If so, then
$1.17\,M_{\odot} < m_c <
1.36\,M_{\odot}$, implying the companion is either a massive white
dwarf or another neutron star.  For the case of a white dwarf
companion, the implication \cite{dc87,ts00a} would be that
PSR~J1906+0746 formed from a binary system of near unity mass ratio in
which both stars were below the critical core-collapse supernova mass
limit $M_{\rm crit} \sim 8 M_{\odot}$. Following a phase in which the
accretion of matter from the evolved and more massive primary star
onto the initially less massive secondary pushed its mass above
$M_{\rm crit}$, the secondary underwent a supernova explosion to form
the currently observable pulsar. 

In the alternative case of a neutron star companion, the small
inferred characteristic age and large magnetic field of PSR~J1906+0746
suggest that PSR~J1906+0746 is the young second-born neutron star,
while the companion is a longer-lived recycled pulsar, spun up to a
period of a few tens of ms during the accretion phase. We have
searched for radio pulsations of this companion but have found none
down to a 1.4-GHz luminosity limit of $\sim$ 0.1~mJy~kpc$^2$; only 0.5\% of
all currently known pulsars \cite{mhth05} have a luminosity below
this value.

These deep searches suggest that the companion to PSR~J1906+0746 is
either: (a) a white dwarf; (b) a faint radio pulsar with a luminosity
below 0.1~mJy~kpc$^2$; or (c) a pulsar whose radio beam does not
intersect our line of sight. Option (a) will be hard to test, as the
$\sim 1$~Myr old white dwarf will have cooled and dimmed. Option (b)
is unlikely to be better testable in the near future (Paper II).

\subsection{System Age and Implications for the Compact-Binary Merger Rate}\label{sec:age}

PSR~J1906+0746 has the smallest characteristic age ($\tau_c=112$~kyr)
of any binary pulsar currently known, but we have not found any
associated supernova remnant (Paper II).

        \begin{figure}[b]%
        \begin{minipage}[b]{84mm}%
          \includegraphics[width=84mm]{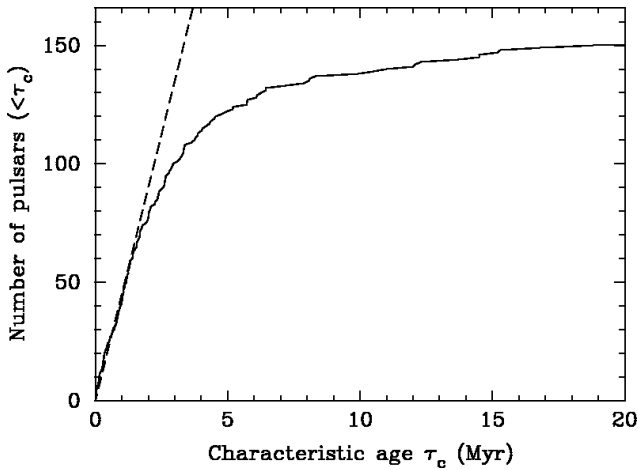}%
        \end{minipage}%
        \hspace{2mm}%
        \begin{minipage}[b]{60mm}%
\caption{\label{fig:cage} 
Cumulative distribution of characteristic ages for pulsars detected
in the PMB with similar surface magnetic fields to PSR~J1906+0746. The
heavy solid line shows the expected trend corresponding to a birthrate
of potentially observable objects of 45~Myr$^{-1}$ (see
text). Luminosity decay and decreasing beaming fractions make older
pulsars harder to detect (Paper II).}
          \vspace{5mm}%
        \end{minipage}%
        \end{figure}%

A simple estimate of the birth rate can be made by considering the
cumulative distribution of characteristic ages. Figure~\ref{fig:cage}
shows this distribution for a sub-sample of 150 pulsars detected in
the PMB which have inferred magnetic field strengths within 1~dB of
PSR~J1906+0746 (i.e.~$|\log(B)-\log(B_{1906})|<0.1$). This sample
shows a linear trend at small ages with a slope of 45~Myr$^{-1}$.
Since the PMB has only detected one J1906+0746-like pulsar in this
sample, the inferred birthrate of {\it similar pulsars potentially
observable in the PMB} is $45/150=0.3$~Myr$^{-1}$. Scaling this rate
to the whole Galaxy (Paper II) we estimate the birth rate of
PSR~J1906+0746-like objects to be $\sim 60 \times (0.2/f)$~Myr$^{-1}$,
where $f$ is the unknown fraction of $4\pi$~sr covered by the radio
beam. For a steady-state population this equals the merger rate, and it
is similar to that derived recently by \cite{kkl+04} for double neutron
star binaries.

The orbital eccentricity of 0.085 for PSR~J1906+0746 is remarkably
similar to that of PSR~J0737$-$3039 \cite{bdp+03}, but smaller than
observed for other double neutron star systems. For PSR~J0737$-$3039
this could be a selection effect as low-eccentricity systems take
longer to coalesce \cite{cb05}.  Only with the detection of further
young systems like PSR~J1906+0746 will we be able to conclude whether
the low eccentricity is a necessary feature of young systems and
how this constrains supernova kicks and pulsar formation.

\subsection{Pulse Profile Evolution}\label{sec:profevol}

Pulsars observed in four other relativistic binaries (PSR~B1913+16:
Weisberg \& Taylor 2003; PSR~B1534+12: Stairs et al.~2004;
PSR~J1141$-$6545: Hotan et al.~2005; PSR~J0737$-$3039B: Burgay et
al.~2005) show mean pulse profile variations with time. The simplest
explanation for this effect is geodetic precession. The profile
variations occur as the precessing pulsar beam changes its orientation
with respect to our line of sight. 

        \begin{figure}[t]%
        \begin{minipage}[b]{40mm}%
\caption{\label{fig:profs} Pulse profiles, integrated for 35 minutes
  at 1.374~GHz for PMB 1998 data (top) and PMB 2005 data (middle).
The lower panel shows
the difference profile after scaling both
profiles to the area of the main pulse. The dashed horizontal lines
show $\pm 3$ standard deviations computed from the off-pulse noise
  region (Paper II).}
          \vspace{5mm}%
        \end{minipage}%
        \hspace{2mm}%
        \begin{minipage}[b]{104mm}%
          \includegraphics[width=104mm]{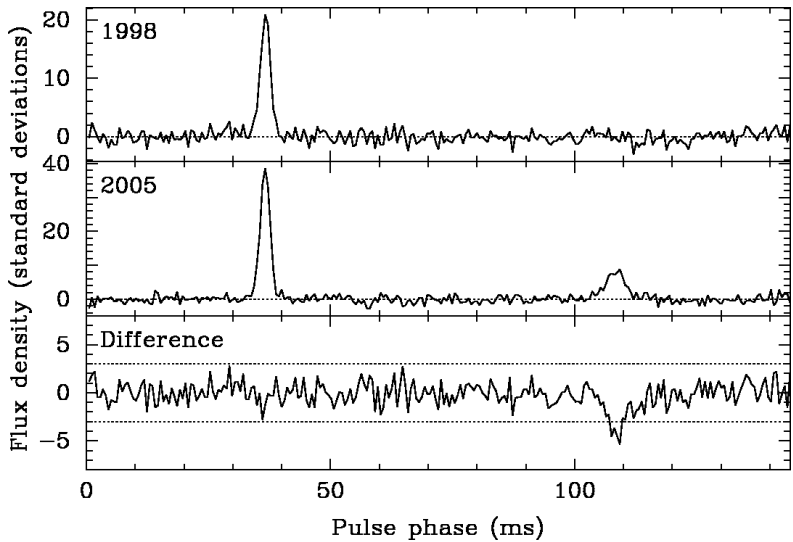}%
        \end{minipage}%
        \end{figure}%

Figure~\ref{fig:profs} shows a first comparison between the integrated
profiles at 1400~MHz from the 1998 PMB detection and a recent Parkes
observation at the same frequency using the same observing system.  We
have scaled each profile to the area of the main pulse and formed the
``difference profile'' by subtracting the 2005 profile from the 1998
one. In the absence of any profile evolution, the difference profile
should be free from systematic trends and have a standard deviation
consistent with the quadratic sum of the off-pulse noise present in
the two input profiles. We observe a significant departure from random
noise around the interpulse region which is not detectable in the 1998
observation (Paper II).  Further observations with better S/N, flux
calibration and polarimetric capability are required to confirm and
quantify these changes.

\subsection{Prospects}
Future radio timing and polarimetric observations of PSR~J1906+0746
should allow the study of several relativistic effects. The very
narrow pulse shape observed means that our
current Arecibo TOAs have an uncertainty of $\sim 5\mu$s in a 5-min
integration. Simulations based on this level of precision show that we
expect to measure the gravitational redshift and time dilation
parameter, $\gamma$, within the coming year and, within a few years,
measure the rate of orbital decay, $\dot{P}_b$. Such measurements
would determine the orbital inclination and masses of the stars so
that the system could possibly be used for further tests of general
relativity. Assuming reasonable ranges for the pulsar mass, the
orbital inclination angle is likely to be in the range
$42^{\circ}<i<51^{\circ}$. For this range of inclinations, a
measurement of the Shapiro delay is less likely at the present level
of precision.  The distance to PSR~J1906+0746 is currently estimated
using the {Cordes} \& {Lazio}~({2002}) electron density model. As for PSR~B1913+16,
kinematic contributions to $\dot{P}_b$, which depend on the assumed
location in the Galaxy and hence the distance, are likely to be a
limiting factor for high-precision tests of general relativity with
this system. We are currently attempting to obtain independent
distance constraints via the detection of neutral hydrogen absorption
and emission.

Comparing pulse profiles taken in 1998 and 2005, we have found some
evidence for long-term evolution of the pulse profile.  Given that the
expected geodetic precession period is only $\sim 200$~yr, the
observed variations could be the first manifestations of this
effect. Future observations with high time resolution and polarimetric
capabilities should provide more quantitative insights.

\section{Conclusions and future work}\label{sec:future}

We have described the initial stages of a large-scale survey for
pulsars using ALFA, the seven-beam system at the Arecibo Observatory
that operates at 1.4 GHz.  Our discovery of 21 new pulsars, including
a young 144-ms pulsar in a highly relativistic 3.98-hr orbit
about a $>0.9~M_{\odot}$ companion, using a preliminary data
acquisition system and analysis is very encouraging.

The full ALFA survey will take more than five years, depending largely
on allocation of telescope time. Numerical models of the pulsar
population, calibrated by results from the PMB survey and
incorporating measured characteristics of the ALFA system, suggest
that as many as 1000 new pulsars will be discovered with the new
spectrometer and the full-resolution analysis. That substantial
increase of the number of known pulsars will not only improve our
understanding of the galactic neutron star population, but also
promises to contain individual systems that probe new regions of
physics.

\acknowledgements 
~ \\ 
This conference presentation was made possible by the Government of
Canada / Cette pr\'esentation a \'et\'e rendu possible avec l'appui du
gouvernement du Canada.\\
Basic research in radio astronomy at the NRL is support by the Office
of Naval Research.


\begin{thebibliography}{}

\bibitem[Abramovici {\rm et~al.}~{1992}]{aad+92}
Abramovici~A. {\rm et~al.}, 1992, Science, 256, 325

\bibitem[Arzoumanian {\rm et~al.}~{2002}]{acc02}
Arzoumanian~Z., Chernoff~D.~F., Cordes~J.~M., 2002, ApJ, 568, 289

\bibitem[Bhattacharya {\rm et al.}~{1992}]{bwhv92} 
Bhattacharya, D., Wijers, R.~A.~M.~J., Hartman, J.~W., \& Verbunt, F.\ 1992, A\&A, 254, 198 

\bibitem[{Burgay} {\rm et~al.}~{2003}]{bdp+03}
{Burgay}~M. {\rm et~al.}, 2003, Nature, 426, 531

\bibitem[Burgay et al.(2005)]{bpm+05} Burgay, M., et al.\ 
2005, ApJ, 624, L113 

\bibitem[Chaurasia \& Bailes~{2005}]{cb05}
Chaurasia~H.~K., Bailes~M., 2005, ApJ, 632, 1054

\bibitem[Clifton {\rm et~al.}~{1992}]{clj+92}
Clifton~T.~R., Lyne~A.~G., Jones~A.~W., McKenna~J., Ashworth~M., 1992, MNRAS,
  254, 177

\bibitem[{Cordes} \& {Lazio}~{2002}]{cl02a}
{Cordes}~J.~M., {Lazio}~T.~J.~W., 2002  {\ttfamily [astro-ph/0207156]}

\bibitem[Cordes {\rm et~al.}~{2005}]{cfl+05}
Cordes~J.~M. {\rm et~al.}, 2005, ApJ, 637, 446

\bibitem[{Demorest} {\rm et~al.}~{2004}]{drb+04}
{Demorest}~P.  {\rm et~al.}, 2004, AAS Abstracts 205

\bibitem[Dewey \& Cordes~{1987}]{dc87}
Dewey~R.~J., Cordes~J.~M., 1987, ApJ, 321, 780

\bibitem[Dowd {\rm et~al.}~{2000}]{dsh00}
Dowd~A., Sisk~W., Hagen~J., 2000, in Kramer~M., Wex~N., Wielebinski~R., eds,
  Pulsar Astronomy - 2000 and Beyond, {IAU} Colloquium 177.
\newblock Astronomical Society of the Pacific, San Francisco, p.~275

\bibitem[Edwards {\rm et al.}~{2001}]{e+01} 
Edwards, R.~T., Bailes, M., van Straten, W., \& Britton, M.~C.\ 2001, \mnras, 326, 358 

\bibitem[{Faulkner} {\rm et~al.}~{2004}]{fsk+04}
{Faulkner}~A.~J. {\rm et~al.}, 2004, MNRAS, 355, 147

\bibitem[Heiles~(2004)]{h04}
Heiles, C. 2004, ALFA Memo, {\tt http://alfa.naic.edu/memos/}

\bibitem[{Hotan {et~al.}~(2005)}]{hbo05}
Hotan, A.~W., Bailes, M., \& Ord, S.~M. 2005, ApJ, 624, 906


\bibitem[Johnston {\rm et~al.}~{1992}]{jlm+92}
Johnston~S., Lyne~A.~G., Manchester~R.~N., Kniffen~D.~A., D'Amico~N., Lim~J.,
  Ashworth~M., 1992, MNRAS, 255, 401

\bibitem[{Kalogera} {\rm et~al.}~{2004}]{kkl+04}
{Kalogera}~V. {\rm et~al.}, 2004, ApJ, 601, L179

\bibitem[Kim {\rm et~al.}~{2004}]{kklw04}
Kim~C., Kalogera~V., Lorimer~D.~R., White~T., 2004, ApJ, 616, 1109

\bibitem[{Lommen} \& {Backer}~{2001}]{lb01}
{Lommen}~A.~N., {Backer}~D.~C., 2001, Bulletin of the American Astronomical
  Society, 33, 1347

\bibitem[{Lorimer} {\rm et~al.}~{2006}]{lsf+06}
{Lorimer}~D.~R. {\rm et~al.}, 2006, ApJ, 640, 428

\bibitem[Lyne {\rm et~al.}~{2004}]{lbk+04}
Lyne~A.~G. {\rm et~al.}, 2004, Science, 303, 1153

\bibitem[Manchester {\rm et~al.}~{2001}]{mlc+01}
Manchester~R.~N. {\rm et~al.}, 2001, MNRAS, 328, 17

\bibitem[{Manchester} {\rm et~al.}~{2005}]{mhth05}
{Manchester}~R.~N., {Hobbs}~G.~B., {Teoh}~A., {Hobbs}~M., 2005, AJ, 129, 1993

\bibitem[Phinney~{1991}]{phi91}
Phinney~E.~S., 1991, ApJ, 380, L17

\bibitem[{Stairs}~{2003}]{sta03}
{Stairs}~I.~H., 2003, Living Reviews in Relativity, 6, 5

\bibitem[{Stairs {et~al.}~(2004)}]{sta04a}
Stairs, I.~H., Thorsett, S.~E., \& Arzoumanian, Z. 2004, PRL, 93,
  141101

\bibitem[{{Tauris} \& {Sennels}~{2000}}]{ts00a}
{Tauris}, T.~M., \& {Sennels}, T. 2000, A\&A, 355, 236


\bibitem[{Weisberg} \& {Taylor}~{2003}]{wt03}
{Weisberg}~J.~M., {Taylor}~J.~H., 2003, in Bailes~M., Nice~D.~J., Thorsett~S.,
  eds, Radio Pulsars.
\newblock Astronomical Society of the Pacific, San Francisco, p.~93

\bibitem[Willems {\rm et~al.}~{2004}]{wkh04}
Willems~B., Kalogera~V., Henninger~M., 2004, ApJ, 616, 414

\end{thebibliography}

\end{document}